\documentstyle[12pt,twoside]{article}

\def\a{\alpha}
\def\b{\beta}

\def\d{\delta}
\def\f{\phi}                    

\def\l{\lambda}
\def\m{\mu}
\def\n{\nu}
\def\o{\omega}
\def\p{\pi}                     
\def\r{\rho}                    
\def\s{\sigma}                  

\def\D{\Delta}

\def\cf{{\cal F}}

\catcode`@=11

\def\un#1{\relax\ifmmode\@@underline#1\else $\@@underline{\hbox{#1}}$\relax\fi}

{\catcode`\'=\active \gdef'{{}~\bgroup\prim@s}}

\def\magstep#1{\ifcase#1 \@m\or 1200\or 1440\or 1728\or 2074\or 2488\or
        2986\fi\relax}

\font\twfvmi=cmmi10\@magscale5
    \skewchar\twfvmi='177
\font\twfvsy=cmsy10\@magscale5
    \skewchar\twfvsy='60
\font\twfvly=lasy10\@magscale5
\font\thtyrm=cmr10\@magscale6

\def\vpt{\textfont\z@\fivrm
  \scriptfont\z@\fivrm \scriptscriptfont\z@\fivrm
\textfont\@ne\fivmi \scriptfont\@ne\fivmi \scriptscriptfont\@ne\fivmi
\textfont\tw@\fivsy \scriptfont\tw@\fivsy \scriptscriptfont\tw@\fivsy
\textfont\thr@@\tenex \scriptfont\thr@@\tenex \scriptscriptfont\thr@@\tenex
\def\prm{\fam\z@\fivrm}%
\def\unboldmath{\everymath{}\everydisplay{}\@nomath
  \unboldmath\fam\@ne\@boldfalse}\@boldfalse
\def\boldmath{\@subfont\boldmath\unboldmath}%
\def\pit{\@getfont\pit\itfam\@vpt{cmti5}}%
\def\psl{\@subfont\sl\it}%
\def\pbf{\@getfont\pbf\bffam\@vpt{cmbx5}}%
\def\ptt{\@subfont\tt\rm}%
\def\psf{\@subfont\sf\rm}%
\def\psc{\@subfont\sc\rm}%
\def\ly{\fam\lyfam\fivly}\textfont\lyfam\fivly
    \scriptfont\lyfam\fivly \scriptscriptfont\lyfam\fivly
\@setstrut\rm}

\def\@vpt{}

\def\vipt{\textfont\z@\sixrm
  \scriptfont\z@\sixrm \scriptscriptfont\z@\sixrm
\textfont\@ne\sixmi \scriptfont\@ne\sixmi \scriptscriptfont\@ne\sixmi
\textfont\tw@\sixsy \scriptfont\tw@\sixsy \scriptscriptfont\tw@\sixsy
\textfont\thr@@\tenex \scriptfont\thr@@\tenex \scriptscriptfont\thr@@\tenex
\def\prm{\fam\z@\sixrm}%
\def\unboldmath{\everymath{}\everydisplay{}\@nomath
  \unboldmath\@boldfalse}\@boldfalse
\def\boldmath{\@subfont\boldmath\unboldmath}%
\def\pit{\@subfont\it\rm}%
\def\psl{\@subfont\sl\rm}%
\def\pbf{\@getfont\pbf\bffam\@vipt{cmbx6}}%
\def\ptt{\@subfont\tt\rm}%
\def\psf{\@subfont\sf\rm}%
\def\psc{\@subfont\sc\rm}%
\def\ly{\fam\lyfam\sixly}\textfont\lyfam\sixly
    \scriptfont\lyfam\sixly \scriptscriptfont\lyfam\sixly
\@setstrut\rm}

\def\@vipt{}

\def\xxxpt{\textfont\z@\thtyrm
  \scriptfont\z@\twfvrm \scriptscriptfont\z@\twtyrm
\textfont\@ne\twfvmi \scriptfont\@ne\twfvmi \scriptscriptfont\@ne\twtymi
\textfont\tw@\twfvsy \scriptfont\tw@\twfvsy \scriptscriptfont\tw@\twtysy
\textfont\thr@@\tenex \scriptfont\thr@@\tenex \scriptscriptfont\thr@@\tenex
\def\unboldmath{\everymath{}\everydisplay{}\@nomath\unboldmath
        \textfont\@ne\twfvmi \textfont\tw@\twfvsy \textfont\lyfam\twfvly
        \@boldfalse}\@boldfalse
\def\boldmath{\@subfont\boldmath\unboldmath}%
\def\prm{\fam\z@\thtyrm}%
\def\pit{\@subfont\it\rm}%
\def\psl{\@subfont\sl\rm}%
\def\pbf{\@getfont\pbf\bffam\@xxxpt{cmbx10\@magscale6}}%
\def\ptt{\@subfont\tt\rm}%
\def\psf{\@subfont\sf\rm}%
\def\psc{\@subfont\sc\rm}%
\def\ly{\fam\lyfam\twfvly}\textfont\lyfam\twfvly
   \scriptfont\lyfam\twfvly \scriptscriptfont\lyfam\twtyly
\@setstrut \rm}

\def\@xxxpt{}

\def\Huge{\@setsize\Huge{36pt}\xxxpt\@xxxpt}

\font\thtymi=cmmi10\@magscale6
    \skewchar\thtymi='177
\font\thtysy=cmsy10\@magscale6
    \skewchar\thtysy='60
\font\thtyly=lasy10\@magscale6
\font\thsirm=cmr12\@magscale6

\def\xxxvipt{\textfont\z@\thsirm
  \scriptfont\z@\thtyrm \scriptscriptfont\z@\twfvrm
\textfont\@ne\thtymi \scriptfont\@ne\thtymi \scriptscriptfont\@ne\twfvmi
\textfont\tw@\thtysy \scriptfont\tw@\thtysy \scriptscriptfont\tw@\twfvsy
\textfont\thr@@\tenex \scriptfont\thr@@\tenex \scriptscriptfont\thr@@\tenex
\def\unboldmath{\everymath{}\everydisplay{}\@nomath\unboldmath
        \textfont\@ne\thtymi \textfont\tw@\thtysy \textfont\lyfam\thtyly
        \@boldfalse}\@boldfalse
\def\boldmath{\@subfont\boldmath\unboldmath}%
\def\prm{\fam\z@\thsirm}%
\def\pit{\@subfont\it\rm}%
\def\psl{\@subfont\sl\rm}%
\def\pbf{\@getfont\pbf\bffam\@xxxpt{cmss12\@magscale6}}%
\def\ptt{\@subfont\tt\rm}%
\def\psf{\@subfont\sf\rm}%
\def\psc{\@subfont\sc\rm}%
\def\ly{\fam\lyfam\thtyly}\textfont\lyfam\thtyly
   \scriptfont\lyfam\thtyly \scriptscriptfont\lyfam\twfvly
\@setstrut \rm}

\def\@xxxvipt{}

\def\HUGE{\@setsize\HUGE{43pt}\xxxvipt\@xxxvipt}

\catcode`@=12

\font\tenex=cmex10 scaled 1200

\def\Sc#1{\hbox{\sc #1}}        

\def\bo{{\raise.05ex\hbox{\large$\Box$}\:}}             
\def\cbo{{\,\raise-.15ex\Sc [\,}}                       
\def\pa{\partial}                                       
\def\su{\sum}                                           
\def\TH{{\raise.2ex\hbox{$\displaystyle \bigodot$}\mskip-4.7mu \llap H \;}}
\def\face{\hbox{\normalsize$\;\;\:{\raise.9ex\hbox{\oo n}\mskip-13mu \llap
        {${\buildrel{\hbox{\frtnrm ..}}\over\smile}$}}\:$}}     
\def\Face{{\raise.2ex\hbox{$\displaystyle \bigodot$}\mskip-2.2mu \llap {$\ddot
        \smile$}}}                                      
\def\Lhat{{\bf\rlap{\kern-.09em$\hat{\phantom L}$}L}}
\def\Lcheck{{\bf\rlap{\kern-.09em$\check{\phantom L}$}L}}

\def\sp#1{{}^{#1}}                              
\def\sb#1{{}_{#1}}                              
\def\sl#1{\rlap{\hbox{$\mskip 1 mu /$}}#1}      
\def\sbra#1{\left\langle #1\right|}             
\def\sket#1{\left| #1\right\rangle}             
\def\svev#1{\left\langle #1\right\rangle}       
\def\leftrightarrowfill{$\mathsurround=0pt \mathord\leftarrow \mkern-6mu
        \cleaders\hbox{$\mkern-2mu \mathord- \mkern-2mu$}\hfill
        \mkern-6mu \mathord\rightarrow$}
\def\dvec#1{\vbox{\ialign{##\crcr
        \leftrightarrowfill\crcr\noalign{\kern-1pt\nointerlineskip}
        $\hfil\displaystyle{#1}\hfil$\crcr}}}           
\def\dt#1{{\buildrel {\hbox{\LARGE .}} \over {#1}}}     
\def\ddt#1{{\buildrel {\hbox{\LARGE .\kern-2pt.}} \over {#1}}}

\def\frac#1#2{{\textstyle{#1\over\vphantom2\smash{\raise.20ex
        \hbox{$\scriptstyle{#2}$}}}}}                   
\def\ha{\frac12}                                        
\def\sfrac#1#2{{\vphantom1\smash{\lower.5ex\hbox{\small$#1$}}\over
        \vphantom1\smash{\raise.4ex\hbox{\small$#2$}}}} 
\def\bfrac#1#2{{\vphantom1\smash{\lower.5ex\hbox{$#1$}}\over
        \vphantom1\smash{\raise.3ex\hbox{$#2$}}}}       
\def\afrac#1#2{{\vphantom1\smash{\lower.5ex\hbox{$#1$}}\over#2}}    

\def\boxes#1{
        \newcount\num
        \num=1
        \newdimen\downsy
        \downsy=-1.64ex
        \mskip-7.8mu
        \bo
        \loop
        \ifnum\num<#1
        \llap{\raise\num\downsy\hbox{$\bo$}}
        \advance\num by1
        \repeat}
\def\boxup#1#2{\newcount\numup
        \numup=#1
        \advance\numup by-1
        \newdimen\upsy
        \upsy=.82ex
        \mskip7.8mu
        \raise\numup\upsy\hbox{$#2$}}

\newskip\humongous \humongous=0pt plus 1000pt minus 1000pt
\def\caja{\mathsurround=0pt}

\newif\ifdtup
\def\panorama{\global\dtuptrue \openup2\jot \caja
        \everycr{\noalign{\ifdtup \global\dtupfalse
        \vskip-\lineskiplimit \vskip\normallineskiplimit
        \else \penalty\interdisplaylinepenalty \fi}}}
\def\li#1{\panorama \tabskip=\humongous                         
        \halign to\displaywidth{\hfil$\displaystyle{##}$
        \tabskip=0pt&$\displaystyle{{}##}$\hfil
        \tabskip=\humongous&\llap{$##$}\tabskip=0pt
        \crcr#1\crcr}}

\def\PL{Phys. Lett. }

\def\PRD{Phys. Rev. D}
\def\ref#1{$\sp{#1]}$}

\parskip=\medskipamount                 
\def\baselinestretch{1.2}       
\thicklines                         
\def\title#1#2#3#4{
\begin{document}
        {\hbox to\hsize{#4 \hfill  #3}}\par
        \begin{center}\vskip.5in minus.1in {\Large\bf #1}\\[.5in minus.2in]{#2}
        \vskip1.4in minus1.2in {\bf ABSTRACT}\\[.1in]\end{center}
        \begin{quotation}\par}
\def\author#1#2{#1\\[.1in]{\it #2}\\[.1in]}

\def\AM{Aleksandar Mikovi\'c\,
\footnote{Supported in part by a grant from Imperial College, London}
\\[.1in] {\it SISSA-International School for Advanced Studies\\
Via Beirut 2-4, Trieste 34100, Italy}\\[.1in]
and \\[.1 in]
{\it Institute for Physics, P.O. Box 57, 11001 Belgrade, Yugoslavia}}

\def\endtitle{\par\end{quotation}\vskip3.5in minus2.3in\newpage}


\def\endabstract{\par\end{quotation}
        \renewcommand{\baselinestretch}{1.2}\small\normalsize}


\def\xpar{\par}                                         
\def\letterhead{
        \centerline{\large\sf IMPERIAL COLLEGE}
        \centerline{\sf Blackett Laboratory}
        \vskip-.07in
        \centerline{\sf Prince Consort Road, SW7 2BZ}
        \rightline{\scriptsize\sf Dr. Aleksandar Mikovi\'c}
        \vskip-.07in
        \rightline{\scriptsize\sf Tel: 071-589-5111/6983}
        \vskip-.07in
        \rightline{\scriptsize\sf E-mail: A.MIKOVIC@IC.AC.UK}}
\def\sig#1{{\leftskip=3.75in\parindent=0in\goodbreak\bigskip{Sincerely yours,}
\nobreak\vskip .7in{#1}\par}}


\def\ree#1#2#3{
        \hfuzz=35pt\hsize=5.5in\textwidth=5.5in
        \begin{document}
        \ttraggedright
        \par
        \noindent Referee report on Manuscript \##1\\
        Title: #2\\
        Authors: #3}


\def\start#1{\pagestyle{myheadings}\begin{document}\thispagestyle{myheadings}
        \setcounter{page}{#1}}


\catcode`@=11

\def\ps@myheadings{\def\@oddhead{\hbox{}\footnotesize\bf\rightmark \hfil
        \thepage}\def\@oddfoot{}\def\@evenhead{\footnotesize\bf
        \thepage\hfil\leftmark\hbox{}}\def\@evenfoot{}
        \def\sectionmark##1{}\def\subsectionmark##1{}
        \topmargin=-.35in\headheight=.17in\headsep=.35in}
\def\ps@acidheadings{\def\@oddhead{\hbox{}\rightmark\hbox{}}
        \def\@oddfoot{\rm\hfil\thepage\hfil}
        \def\@evenhead{\hbox{}\leftmark\hbox{}}\let\@evenfoot\@oddfoot
        \def\sectionmark##1{}\def\subsectionmark##1{}
        \topmargin=-.35in\headheight=.17in\headsep=.35in}

\catcode`@=12

\def\sect#1{\bigskip\medskip\goodbreak\noindent{\large\bf{#1}}\par\nobreak
        \medskip\markright{#1}}
\def\chsc#1#2{\phantom m\vskip.5in\noindent{\LARGE\bf{#1}}\par\vskip.75in
        \noindent{\large\bf{#2}}\par\medskip\markboth{#1}{#2}}
\def\Chsc#1#2#3#4{\phantom m\vskip.5in\noindent\halign{\LARGE\bf##&
        \LARGE\bf##\hfil\cr{#1}&{#2}\cr\noalign{\vskip8pt}&{#3}\cr}\par\vskip
        .75in\noindent{\large\bf{#4}}\par\medskip\markboth{{#1}{#2}{#3}}{#4}}
\def\chap#1{\phantom m\vskip.5in\noindent{\LARGE\bf{#1}}\par\vskip.75in
        \markboth{#1}{#1}}
\def\refs{\bigskip\medskip\goodbreak\noindent{\large\bf{REFERENCES}}\par
        \nobreak\bigskip\markboth{REFERENCES}{REFERENCES}
        \frenchspacing \parskip=0pt \renewcommand{\baselinestretch}{1}\small}
\def\unrefs{\normalsize \nonfrenchspacing \parskip=medskipamount}
\def\Item{\par\hang\textindent}
\def\Itemitem{\par\indent \hangindent2\parindent \textindent}
\def\makelabel#1{\hfil #1}
\def\topic{\par\noindent \hangafter1 \hangindent20pt}
\def\Topic{\par\noindent \hangafter1 \hangindent60pt}

\title{Hawking Radiation and Back-Reaction in a Unitary Theory of 2D
Quantum Gravity}
{\AM}{98/94/EP}{July 1994}
We investigate the semiclassical limit and quantum corrections to the metric
in a unitary quantum gravity formulation of the CGHS 2d dilaton gravity model.
A new method for calculating the back-reaction effects has been introduced,
as an expansion of the effective metric in powers of the matter energy-momentum
tensor. In the semiclassical limit the quantum corrections can be neglected,
and
we show that physical states exists which contain the Hawking
radiation. The first order back-reaction effect
is entirely due to the Hawking radiation. It causes the black-hole mass
to monotonically decrease, and it makes it unbounded from bellow as the horizon
is approched. The second order quantum corrections have been estimated.
Since the matter is propagating freely in this unitary theory, we
expect that the higher order corrections
will stabilize the mass, and the black hole will completely evaporate leaving
a nearly flat space.

\endtitle

\noindent The question of the back-reaction of the Hawking radiation is
important for understanding the quantum fate of a black hole. Since a
proper treatment of the problem requires a quantum gravity theory, not much
progress has been made in 4d. However, in 2d, the CGHS dilaton gravity theory
represents an excellent toy model for study of the backreaction problem
\cite{cghs}. Moreover, a quantum gravity formulation
exists \cite{{mik1},{mik2}}, and the model can be thought of as
a crude approximation of the BCMN model, which describes a
spherically symmetric scalar field collapse in 4d \cite{{bcmn},{unru}}.
Hence, the CGHS model
offers a possibility to study the backreaction problem in a quantum gravity
context. Another interesting point is that the quantum gravity formulation
of CGHS can be made unitary \cite{{mik1},{mik2}}. The reason is that the
theory can be deparametrized, and the corresponding Hamiltonian can be
promoted into a Hermitian operator, because it is just a free-field
Hamiltonian.
Note that Hajicek has used exactly the same reasoning
some time ago to argue that the
quantum theory of the BCMN model could be made unitary \cite{haj}.
However, the problem
there is that the reduced Hamiltonian is a complicated nonlocal function of
the matter fields, and hence the existence of a Hermitian extension is not
guaranteed. Also the BCMN gauge does not penetrate the horizon, and hence a
different slicing, which should be complete, is nedeed.

Therefore one can explore the problem of whether the evaporating black holes
can exist within a unitary quantum gravity theory in 2d.
In \cite{mik2}, a somewhat vague arguments were
given in the support of the claim that the Hawking radiation exists in the
semiclassical limit of the 2d quantum dilaton gravity. An argument of
a different type has also been made in \cite{verl}. In this paper we present
concrete calculations supporting the argument given in \cite{mik2} for
the existence of the Hawking radiation. Moreover, we present a method for
calculating the backreaction
corrections to the metric, which follows from the solvability of the
classical theory. The metric is known as an explicit function of the matter
energy-momentum tensor $T_m$. The corresponding nonlocal operator is then
appropriately expanded into powers of the
$T_m$ operator, and the effective quantum metric is calculated as an
expectation value of the metric operator in a physical state.
In the zeroth
order one obtains the classical metric corresponding to the classical matter
content of the quantum state, while the n-th order correction is
proportional to the expectation value of $(T_m)^n$ in the initial quantum
state.

We start from the CGHS action \cite{cghs}
$$ S =  \int_{M} d^2 x \sqrt{-g} \left[ e^{-\f}\left( R +
 (\nabla \f)^2 + 4\l^2 \right) - \su_{i=1}^N (\nabla f_i)^2 \right]\quad,
\eqno(1)$$
where $\f$ is a dilaton, $f_i$ are scalar matter fields,  $g$, $R$
and $\nabla$ are
determinant, scalar curvature and covariant derivative respectively,
associated with a metric $g_{\m\n}$ on the 2d manifold $M$. Topology of
$M$ is that of $ {\bf R} \times {\bf R}$. We make a field redefinition
$$ \tilde{g}_{\m\n} = e^{-\f} g_{\m\n} \quad,\quad \tilde{f} = e^{-\f} \quad,
\eqno(2)$$
so that
$$ S =  \int_{M} d^2 x \sqrt{-\tilde{g}}\left[
 \tilde{R}\tilde{f} + 4\l^2 - \su_{i=1}^N (\tilde{\nabla} f_i)^2 \right]\quad.
 \eqno(3)$$
Canonical analysis of (3) \cite{mik2} gives
$$ S= \int dt dx \left( \p_{\tilde\r}\dt{\tilde\r} +
\p_{\tilde\f}\dt{\tilde\f} + \p_{f}\dt{f} -
 N_0 G_0 - N_1 G\sb 1 \right) \quad, \eqno(4)$$
where $N_0$ and $N_1$ are the laps and shift, while
the constraints $G_0$ and $G_1$ are given as
$$\li{G\sb 0  =& - \p_{\tilde\r} \p_{\tilde\f}  - 4\l^2 e^{\tilde\r} +
2\tilde{\f}^{\prime\prime} - \tilde{\r}^{\prime}\tilde{\f}^{\prime} +
 \ha (\p_f^2 + (f^{\prime})^2)\cr
G\sb 1  =& \p_{\tilde{\f}} {\tilde{\f}}^{\prime} + \p_{\tilde{\r}}
\tilde{\r}^{\prime} -2 \p_{\tilde{\r}}^{\prime}
+ \p_f f^{\prime} \quad.&(5)\cr} $$
The primes stand for the $x$ derivatives, $\tilde\r$
is the conformal factor ($\tilde{g} = e^{\tilde\r}$), and we have taken $N=1$
for the simplycity sake.
Now we fix the gauge
$$ \tilde\r = 0 \quad,\quad \p_{\tilde\f }= 0 \quad,\eqno(6)$$
which can be thoutght of as the 2d dilaton gravity analog of the BCMN gauge.
However, unlike the BCMN gauge, (6) penetrates the horizon and corresponds
to the classical CGHS solutions where the gauge functions are fixed
to be $x^{\pm} = t \pm x $ \cite{mik2}.
Solving the constraints gives
$$\tilde\f = a + bx + \l^2 x^2 - \frac14 \int dx\int dx (\p_f^2 +
(f^{\prime})^2) \quad,\quad \p_\r = c + \ha\int dx \p_f f^{\prime}
\quad, \eqno(7)$$
so that the independent canonical variables (or true degrees of freedom) are
$(\p_f, f)$ canonical pairs. The reduced phase space Hamiltonian is a
free-field
Hamiltonian
$$ H = \ha\int_{-\infty}^{\infty}dx\,(\p_f^2 +
(f^{\prime})^2)\quad. \eqno(8)$$
The dilaton and the original metric can be expressed in the gauge (6) as
$$e^{-\f} =  - \l^2 x^+ x^- - F_+ - F_- \quad,\quad ds^2 = -e^{\f}dx^+ dx^-
\quad,\eqno(9)$$
where
$$ F_{\pm}= \int_{-\infty}^{\infty} dy G(x^{\pm} -y) T_{\pm\pm} (y)
\quad.\eqno(10)$$
$G$ is a 2d Green's function ($G^{\prime\prime}(x)=\d(x)$), and we take
$G(x)=x\theta (x)$, where $\theta(x)$ is a step function. $T_{\pm\pm}$
is the matter energy-momentum tensor
$$T_{\pm\pm} = \ha \pa_{\pm} f \pa_{\pm} f \quad.\eqno(11)$$
The formulas (9-10) can be derived from (7) by using $\p_f = \dt{f}$.

Quantum theory is defined by choosing a representation of the quantum
canonical commutation relations
$$ [\p_f (x), f(y) ] = -i \d(x-y) \quad.\eqno(12)$$
We take the standard Fock space representation, by defining the creation and
anhilation operators $a^{\dagger},a$ as
$$ a_k = {-i \p_k + k\,{\rm sign}(k) f_k\over\sqrt{2|k|}} \quad,\eqno(13)$$
where
$$f(x)=\int_{-\infty}^{\infty}{dk\over\sqrt{2\p}}e^{ikx} f_k \quad,\quad
\p_f (x)=\int_{-\infty}^{\infty}{dk\over\sqrt{2\p}}e^{ikx}\p_k
\quad,\eqno(14)$$
so that (12) is equivalent to
$$ [a_k , a_q^{\dagger} ] = \d (k-q) \quad.\eqno(15)$$
The Fock space ${\cf} (a_k)$ with the vacuum $\sket{0}$ is the physical
Hilbert space of the theory. The
Hamiltonian (8) can be promoted into a Hermitian operator acting on ${\cf}$
as
$$H = \int_{\infty}^{\infty} dk\, \o_k a_{k}^{\dagger} a_k + E_0 \eqno(16)$$
where $\o_k = |k|$ and $E_0$ is the vacuum energy.
Therefore one has a unitary evolution described by a Schr\"odinger
equation
$$ i{\pa\over\pa t} \Psi (t) = H \Psi(t) \quad,\eqno(17) $$
where $\Psi(t) \in \cf$.
It will be convenient to work in the Heisenberg picture
$$ \Psi_0 = e^{iHt}\Psi(t) \quad,\quad A (t) =
e^{iHt}Ae^{-iHt}\quad,\eqno(18)$$
so that
$$f(t,x) = e^{iHt} f(x) e^{-iHt} = {1\over\sqrt{2\p}}\int_{-\infty}^{\infty}
{dk\over\sqrt{2\o_k}}\left[ a_k e^{i(kx-\o_k t)} + a_k^{\dagger}
e^{-i(kx-\o_k t)}\right]\quad.\eqno(19)$$
It is also useful to split (19) into left and right moving parts, so that
$f = f_+ + f_-$ where
$$f_{\pm} (x^{\pm}) = {1\over\sqrt{2\p}}\int_{0}^{\infty}
{dk\over\sqrt{2\o_k}}\left[ a_{\pm}(k) e^{-ikx^{\pm}} + a_{\pm}(k)^{\dagger}
e^{ikx^{\pm}}\right]\quad.\eqno(20)$$
The metric is given by the operator $e^\f$, which can be defined as
the inverse of the
$e^{-\f}$ operator. The $e^{-\f}$ operator in the Heisenberg picture
can be easily defined from the expressions (9) and (10), where now $f$ is the
operator given by (19), while in the expressions for $T_{\pm\pm}$ there will
be a normal ordering with respect to some vacuum in $\cf$, which can be
different from $\sket{0}$.

Given a physical state $\Psi_0$, one can associate an effective metric to
$\Psi (t)$ as
$$e^{\r_{eff} (t,x)} = \sbra{\Psi (t)}e^{\f (x)}\sket{\Psi (t)}
=\sbra{\Psi_0}e^{\f (t,x)}\sket{\Psi_0} \quad.\eqno(21)$$
$e^{\r_{eff}}$ can be interpreted as a metric if $\svev{(e^\f)^2}-
\svev{e^\f}^2$
is sufficiently small. This deviation cannot be zero, since if it was zero
it would mean that $\Psi_0$ is
an eigenvalue of the metric, whose spectrum is continious while $\Psi_0$ is a
normalisable state.
In order to calculate (21), we use the following formal identity
$$(-\l^2 x^+ x^- -  F )^{-1} = e^{\f_0}(1 - e^{\f_0}\d F)^{-1} =
e^{\f_0} \su_{n=0}^{\infty} e^{n\f_0} \d F^n \quad,\eqno(22)$$
where $F_0$ is a c-number function, $e^{-\f_0} = -\l^2 x^+ x^- -  F_0 $ and
$\d F = F - F_0$. Then
$$\svev{(-\l^2 x^+ x^- -  F )^{-1}} =
e^{\f_0} \su_{n=0}^{\infty} e^{n\f_0} \svev{\d F^n}\quad.\eqno(23)$$

We now want to choose $\Psi_0$ such that it is as close as possible to the
classical matter distribution $f_0 (x^+)$ describing  a left-moving pulse of
matter. The corresponding classical metric is described by
$$ e^{-\r_0} = {M(x^+)\over \l} - \l^2 x^+ \D (x^+) - \l^2 x^+ x^- \eqno(24)$$
where
$$ M (x^+) = \l\int_{-\infty}^{x^+} dy\, y \,T_{++}^0 (y)\quad,\quad
\l^2 \D = \int_{-\infty}^{x^+} dy\, T^0_{++} (y) \quad \eqno(25)$$
and $T_{++}^0 = \ha \pa_{+}f_0 \pa_{+} f_0$. The geometry is that of the black
hole of the mas $M = {\rm lim}_{x^+ \to +\infty} M(x^+)$ and the horizon is at
$ x^- = -\D = -{\rm lim}_{x^+ \to +\infty} \D (x^+)$. The asymptotically flat
coordinates $(y^+,y^-)$ at the past infinity  are given by \cite{gn}
$$ \l x^+ = e^{\l y^+} \quad,\quad \l x^- = - e^{-\l y^-} \quad,\eqno(26)$$
while the asymptotically flat coordinates $(\s^+,\s^-)$ at the future infinity
satisfy
$$ \l x^+ = e^{\l \s^+} \quad,\quad \l (x^- + \D ) = - e^{-\l \s^-}\quad.
\eqno(27)$$

Since the particle vacuum for the observer at the past infinity
is defined with respect to the $y^{\pm}$ coordinates, we will introduce the
in creation and anhilation operators $a_{in}^{\dagger},a_{in}$ as
$$f_+ (t,x) = {1\over\sqrt{2\p}}\int_{0}^{\infty}
{dk\over\sqrt{2\o_k}}\left[ a_{in}(k) e^{-iky^+} + a_{in}(k)^{\dagger}
e^{iky^+}\right]\quad,\eqno(28)$$
and similarly for the right-moving sector. Hence we take for $\Psi_0$ a
coherent state
$$\Psi_0 = e^A \sket{0_{in}^+}\otimes \sket{0_{in}^-} \eqno(29)$$
where
$$A = \int_0^{\infty}dk [f_0 (k) a_{in}^{\dagger}(k) - f_0^{*}(k) a_{in}(k)]
\eqno(30)$$
and $f_0 (k)$ are Fourier modes of $f_0 (y^+)$. Before we start computing
the effective metric, we have to choose the function $F_0$, and we have
to fix the
normal ordering in the $F$ operator. We take $F_0$ corresponding to the
classical metric (24), since we want to compute the quantum corrections to it,
and we take the normal ordering with respect to the out vacuum. From (26-27)
follows
that $y^+ = \s^+$ so that $\sket{0_{in}^+}=\sket{0_{out}^+}$ and hence in the
left sector ``out" normal ordering is the same as the ``in" normal ordering.
This is not the case
in the right sector, where the in and out vacuum are related by
$$ \sket{0_{in}^-}= S(a_{out},a_{out}^{\dagger})\sket{0_{out}^-}\eqno(31)$$
and $S$ is the generator of the Bogoliubov transformation \cite{gn}
$$ a_{in}(k) = S a_{out}(k) S^{\dagger} =
\int_0^{\infty} dq\, ( b_q \a_{qk} + b_q^{\dagger} \b_{qk}^{*} +
\hat{b}_q \hat{\a}_{qk} + \hat{b}_q^{\dagger} \hat{\b}_{qk}^{*})
\quad,\eqno(32)$$
where $a_{out}$ are split into operators outside and inside the horizon,
represented by $b$ and $\hat{b}$, respectively.

Now it is not difficult to show that $\svev{F_+} = F_0$, while
$$ \svev{F_-} = \int_{-\infty}^{x^-} dx_1^- (x^- - x_1^- )
\svev{T_{--}(x_1^-)} \quad.\eqno(33)$$
Note that
$$\svev{T_{--}(x^-)}=\left({\pa\s^-\over\pa x^-}\right)^2
\svev{T_{--}(\s^-)}\quad,\eqno(34)$$
and from \cite{{cghs},{gn}} we have
$$ \svev{T_{--}(\s^-)} = {\l^2\over 48}\left[ 1 - (1 +
\l\D e^{\l\s^-})^{-2}\right]\quad,\eqno(35)$$
so that
$$\svev{F_-} = \int_{-\infty}^{x^-} dx_1 (x^- - x_1 ) {1\over 48}
( (x_1 +\D)^{-2} - x_1^{-2})
         = - {1\over 48}\log\Big|{x^- + \D\over x^-}\Big|\quad.\eqno(36)$$
This gives
$$e^{\r_{eff}} \approx e^{\r_0} ( 1 + e^{\r_0}\svev{F_-} )\quad,\eqno(37)$$
which is a good approximation if
$$ |e^{\r_0}\svev{F_-}|={{1\over 48}\log|{x^- +\D\over x^-}|
\over |{M\over \l} - \l^2 x^+ (\D + x^-)|} << 1 \quad.\eqno(38)$$
Eq. (38) implies
$$  {\log|\l\D|\over 48} + {\l\s^-\over 48} << {M\over\l} + \l x^+ e^{-\l\s^-}
\quad, \eqno(39)$$
so that if ${M\over\l}>>{1+\log|\l\D|\over 48}$
then for a considerable time we can neglect the quantum corrections, and
we effectively have a quantum field propagation on a black hole background.
As is well known, this will produce the Hawking radiation, with the
temperature $T_H ={\l\over 2\p}$ and the flux at
the future infinity given by (35) \cite{cghs}. Hence the Hawking
radiation is present in
the semiclassical limit of our unitary quantum gravity theory.

Note that when $|e^{\r_0}\svev{F_-}|<< 1$, then
$$e^{\r_{eff}} \approx e^{\r_0} ( 1 + e^{\r_0}\svev{F_-} )
 \approx e^{\r_0} ( 1 - e^{\r_0}\svev{F_-} )^{-1} = (e^{-\r_0}
-\svev{F_-})^{-1}
\quad.\eqno(40)$$
The last expression in eq. (40) can be interpreted as the semiclassical
metric with the
first order backreaction effect included. It can be written as
$$ e^{-\r_1} = {M(x^+)\over \l} - \l^2 x^+ \D (x^+) - \l^2 x^+ x^-
+ {1\over 48}\log\Big|{x^- + \D\over x^-}\Big|\quad.\eqno(41)$$
As expected, the curvature singularity is still present at this order, but
what is interesting is the behaviour of a local mass function, which can
be defined as
$$ M_1 (x^+,x^-) ={1\over 4\l} e^{-\f_1} R_1 \quad.\eqno(42)$$
One then obtains
$$ M_1 (x^+,x^-) = M (x^+ ) + {\l\over 48}\log\Big| 1 + {\D\over x^-}\Big| +
{\l\over 48}{\D\over x^-} \quad,\eqno(43)$$
which grows from $0$ to $M$ for large negative $x^-$ (corresponding to
the formation of the black hole at early times), but then it starts to
monotonically decrease, and it reaches zero at some $x_0^- < -\D$. As
$x^-$ is approaching the horizon
$-\D$ (or for very late late times), $M_1$ goes to $-\infty$.
Hence in the region of
validity of our approximation ($x^-$ not to close to $-\D$, which is equivalent
to not too late times) the backreaction effect is precisely what one expects,
i.e. black hole mass decreases due to the Hawking radiation. As time approaches
the future infinity, one expects that the higher order corrections will
stabilize the mass. One can think of various possibilities for this
stabilized mass function, but the one which fits most naturally with the fact
that the matter is freely propagating, is that mass goes to zero, i.e. black
hole
completely evaporates end one ends up with a nearly flat space-time. Since the
theory is unitary, the purity of the state is preserved through non-local
quantum correlations \cite{mik2}.

In order to see whether something like this will happen, we will have to
calculate the higher order corrections. Note that if we use $F_0 = \svev{F_+}
+ \svev{F_-}$, the zero order metric will be the semiclassical metric
$e^{\r_1}$, and the expansion (23) will become more symmetric, with $n=1$ term
vanishing. Calculating the $\svev{\d F^n}$ terms will require calculating
$ \svev{ T(x_1) \cdots T(x_2) } $ and this will require a regularization.
The singularity structure of such an expression is encoded in the OPE
$$ T (x) T (y) = {c/2\over (x - y)^4} +
{2T (y)\over (x - y)^2} + {2\pa T (y)\over x - y }
+ {\rm const.} + o(x - y)\quad, \eqno(44)$$
and the simplest thing one can do is to define
$$:T(x_1) \cdots T(x_n): = T(x_1) \cdots T(x_n) -\sbra{0}T(x_1) \cdots T(x_n)
\sket{0} \quad,\eqno(45)$$
where $\sket{0}$ is the relevant vacuum (i.e. $\sbra{0}T\sket{0} =0$).

In the $n=2$ case we will have
$$\svev{:T(x_1) T(x_2):} = \sbra{0}e^{-X}:T(x_1) T(x_2):e^X \sket{0}
=\sbra{0}:T_X (x_1) T_X (x_2): \sket{0} \eqno(46)$$
where $T_X = e^{-X}Te^X$ and $X=A$ for the left sector, while $X=\log S$ for
the right sector. Then by using (44) and (45) we get
$$\sbra{0}:T_X (x_1) T_X (x_2): \sket{0} =
{2\sbra{0} T_X (x_2) \sket{0}\over (x_1 - x_2 )^2} +
{2\pa \sbra{0}T_X (x_2) \sket{0}\over x_1 - x_2 } +
o(x_1 - x_2)\quad, \eqno(47)$$
and since $\sbra{0} T_X \sket{0} \ne 0$, the singularity is still
present. However, in the left sector, one can calculate exactly (47)
by using $T_X = T-[X,T] + \ha [X,[X,T]] + \cdots$, where
it terminates after the third term, and one obtains
$$\li{\sbra{0}:T_A (y_1) T_A &(y_2): \sket{0} - T_0 (y_1) T_0 (y_2)=
\sbra{0} [A,T])_1 [A,T])_2\sket{0}\cr
=& -\int_0^{\infty} dk\,k\, e^{ik(y_1 -y_2 )}
{\pa f_0\over\pa y_1}{\pa f_0\over\pa y_2} \quad.&(48)\cr}$$
Since
$$ \int_0^{\infty} dk\,k\, e^{ik(y_1 - y_2 )} =
-{1\over (y_1 - y_2 )^2} \quad,\eqno(49)$$
one can get back to the OPE form, but we will not do that. Instead we write
$$\li{\svev{\d F_+^2} =& - \prod_{i=1}^2 \int_{-\infty}^{y^+} dy_i
(e^{\l(y^+ - y_i)} - 1) \int_0^{\infty} dk\,k\, e^{ik(y_1 -y_2 )}
{\pa f_0\over\pa y_1}{\pa f_0\over\pa y_2}\cr =&
\int_0^{\infty} dk\,k\, |\cf (k, y^+)|^2 \quad,&(50)\cr}$$
where
$$\cf (k,y^+) = \int_{-\infty}^{y^+} dy e^{iky}(e^{\l(y^+ - y)} - 1)
{\pa f_0\over\pa y} \quad.\eqno(51)$$
Then by choosing $f_0$ which falls quickly enough away from the centre of the
matter pulse, we can get $\cf (k)$ such that (49) is finite.

In the right sector one can apply a similar trick, although the actual
calculation is more difficult. Calculating
$\svev{T_{--}(1) \cdots T_{--}(n)}$ requires  calculating
$$ \sbra{0_{in}^-} a_{out}^{\pm} (1) \cdots a_{out}^{\pm} (2n) \sket{0_{in}^-}
\quad,\eqno(52)$$
where $a_{out}^{\pm}$ stands for the creation or the anhilation operator,
respectively. Expression (52)
boils down to a sum of products of the following integrals
$$ \int_0^{\infty} dp\, \b_{kp}^{\pm} \b_{pq}^{\pm}\quad,\quad
\int_0^{\infty} dp\, \b_{kp}^{\pm} \a_{pq}^{\pm}\quad,\quad
\int_0^{\infty} dp\, \a_{kp}^{\pm} \a_{pq}^{\pm}\quad,\eqno(53)$$
where $X^{\pm}= X$ or $X^*$, and $\a$ and $\b$ are the coefficients of the
Bogoliubov transformation (32). One can calculate the integrals in (53)
for late times, since then
$$ \int_0^{\infty} dp\, \b_{kp}^{*} \b_{pq} \approx \svev{n_k}\d (k-q) =
{e^{-\b k}\over 1 - e^{-\b k}} \d (k-q) \eqno(54)$$
and
$$ \a_{kq} \approx -e^{\b k/2} \b_{kq}\quad,\eqno(55) $$
where $\b={2\p\over\l}$ \cite{gn}, and consequently obtain
a good estimate of (52). From (54) one can expect that the
convergence will come from the exponentaly damping factor $e^{-\b k}$ in
$\svev{n_k}$. For example
$$\li{\svev{F_-^2} \approx {1\over\l^2}\prod_{i=1}^2 & \int_{-\infty}^{\s^-}
d\s_i  (e^{-\l(\s^- - \s_i)} - 1 )\Big[
{c_0\over (\s_1 - \s_2)^4} + \cr &
{c_1\over (\s_1 - \s_2)^2} {\rm sh}^{-2}(\l(\s_1 - \s_2)/2)
  + c_2 {\rm ch}^{-4}(\l(\s_1 - \s_2)/2)\Big]
\quad,&(56)\cr}$$
where $c_i$ are certain numerical constants. Note that $c_0 +{4\over\l^2} c_1
=0$ because of the absence of $(\s - \s_i)^{-4}$ term in the OPE. However,
there is still a divergence due to $(\s -\s_i)^{-2}$ terms, and hence we use
eq. (49) to rewrite the first two terms in (56) as
$$\li{-{1\over\l^2}\int_0^{\infty} dk\,k\,
 \prod_{i=1}^2 \int_{-\infty}^{\s^-}
d\s_i  & (e^{-\l(\s^- - \s_i)} - 1 ) e^{ik(\s_1 - \s_2)}\cr &\Big[
-{1\over (\s_1 - \s_2)^2} +  {\l^2 \over 4}{\rm sh}^{-2}(\l(\s_1 - \s_2)/2)
\Big] \quad,&(57)\cr}$$
which gives a finite expression.

In conclussion, we have find a way to incorporate the Hawking radiation and
backreaction effects in a manifestly unitary 2d quantum gravity theory.
It still remains to be investigated how to regulate the higher order
corrections, and whether a cut-off in momentum will be needed. The cut-off
would
require a renormalization scheme, and since the CGHS theory is renormalisable,
the hope is that will symplify the analysis in our case. Related to that is the
question of the sumability of the perturbative series in order to obtain
the exact quantum metric. Another possibility
is to define the effective quantum metric recursivelly as
$$ e^{\r_{n+1}} = e^{\r_n} ( 1 + e^{\r_n}\svev{F- F_n } +
e^{2\r_n}\svev{(F- F_n )^2} ) =(-\l^2 x^+ x^- - F_{n+1} )^{-1}\eqno(58)$$
starting from $n=1$. One would then obtain a series of geometries $G_n$,
and the question then would be to find a meanningful limit of such a series.
As far as the 4d black holes are concerned, our method could be applied to
the BCMN model. However, a modification will be neccessary, since the
BCMN model is not classically solvable, and hence one will not have
an explicit expression for the metric as a function of the matter fields,
which is the basis of our approach.

\end{document}